\documentclass[amssymb,preprint,aps,superscriptaddress]{revtex4}
\usepackage{amsmath}
\usepackage{graphicx}
\usepackage{verbatim}
\usepackage{graphics}
\usepackage{mathdots}
\usepackage{color}
\usepackage{ulem}

\newcommand{\subfig}[2]{Fig.~\ref{fig:#1}(#2)}

\begin{document}

\title{Bipolar spin blockade and coherent state\\ superpositions in a triple quantum dot}
\author{M. Busl}
\affiliation{Instituto de Ciencia de Materiales de Madrid, CSIC, Cantoblanco, 28049 Madrid, Spain}

\author{G. Granger}
\affiliation{National Research Council Canada, 1200 Montreal Rd., Ottawa, ON K1A 0R6 Canada}

\author{L. Gaudreau}
\affiliation{National Research Council Canada, 1200 Montreal Rd., Ottawa, ON K1A 0R6 Canada}
	\affiliation{D\'epartement de physique, Universit\'e de Sherbrooke, Sherbrooke, QC J1K 2R1 Canada}

\author{R. S\'anchez}
\affiliation{Instituto de Ciencia de Materiales de Madrid, CSIC, Cantoblanco, 28049 Madrid, Spain}

\author{A. Kam}
\affiliation{National Research Council Canada, 1200 Montreal Rd., Ottawa, ON K1A 0R6 Canada}

\author{M. Pioro-Ladri\`ere}
\affiliation{D\'epartement de physique, Universit\'e de Sherbrooke, Sherbrooke, QC J1K 2R1 Canada}

\author{S.~A.~Studenikin}
\affiliation{National Research Council Canada, 1200 Montreal Rd., Ottawa, ON K1A 0R6 Canada}

\author{P. Zawadzki}
\affiliation{National Research Council Canada, 1200 Montreal Rd., Ottawa, ON K1A 0R6 Canada}

\author{Z. R. Wasilewski}
\affiliation{National Research Council Canada, 1200 Montreal Rd., Ottawa, ON K1A 0R6 Canada}

\author{A. S. Sachrajda}
\affiliation{National Research Council Canada, 1200 Montreal Rd., Ottawa, ON K1A 0R6 Canada}

\author{G. Platero}
\email{gplatero@icmm.csic.es}
\affiliation{Instituto de Ciencia de Materiales de Madrid, CSIC, Cantoblanco, 28049 Madrid, Spain}

\date{\today}
\maketitle
{\bf
Spin qubits based on interacting spins in double quantum dots have been successfully demonstrated~\cite{pettaSC05,hansonPRL07}. Readout of the qubit state involves a conversion of spin to charge information, universally achieved by taking 
advantage of a spin blockade phenomenon resulting from Pauli's exclusion principle. The archetypal 
spin blockade transport signature in double quantum dots takes the form of a rectified current~\cite{onoSC02}. Currently 
more complex spin qubit circuits including triple quantum dots are being developed~\cite{gaudreauNAPHYS12}. Here we show both 
experimentally and theoretically (a) that in a linear triple quantum dot circuit, the spin blockade becomes bipolar \cite{hsieh2012} with current strongly suppressed in both bias directions and (b) that a new quantum coherent mechanism becomes relevant. Within this mechanism
charge is transferred 
non-intuitively via coherent states from one end of the linear triple dot circuit to the other without involving the centre site. Our results have implications in future complex nano-spintronic circuits.
}

Coherent coupling of quantum states can lead to molecular-like superpositions where the electronic wave function has no weight in a spatial region of the system. A simple example of this is the anti-bonding orbital of the $\text{H}_2^+$ molecule. In an array of serially coupled states, coherent superpositions can be formed that are found at the two extremes of the chain avoiding the occupation of the intermediate states. These kinds of orbitals have been shown to be responsible for dark resonances in the fluorescence of sodium atoms~\cite{arimondoCPT}. 
Electronic transport through quantum dot arrays has been predicted to be similarly affected by dark states~\cite{brandesPRL00}. In a triple quantum dot (TQD) with a triangular arrangement, a dark state is predicted to switch off the current flow~\cite{emaryEPL06}. An alternative geometry, where the source is connected to one outer dot and the drain to the other, involves a resonance at the two ends of the chain leading to transport regardless of the configuration of the intermediate site~\cite{ratnerJPhysChem90}. These effects are predicted to enable new functionalities such as adiabatic passage or quantum rectification~\cite{emaryEPL06, Greentree2004}. In this paper we show that equivalent superpositions manifest themselves as a resonant leakage current through a spin blockaded TQD.

\begin{figure}
\begin{center}
\includegraphics[width=5in,clip]{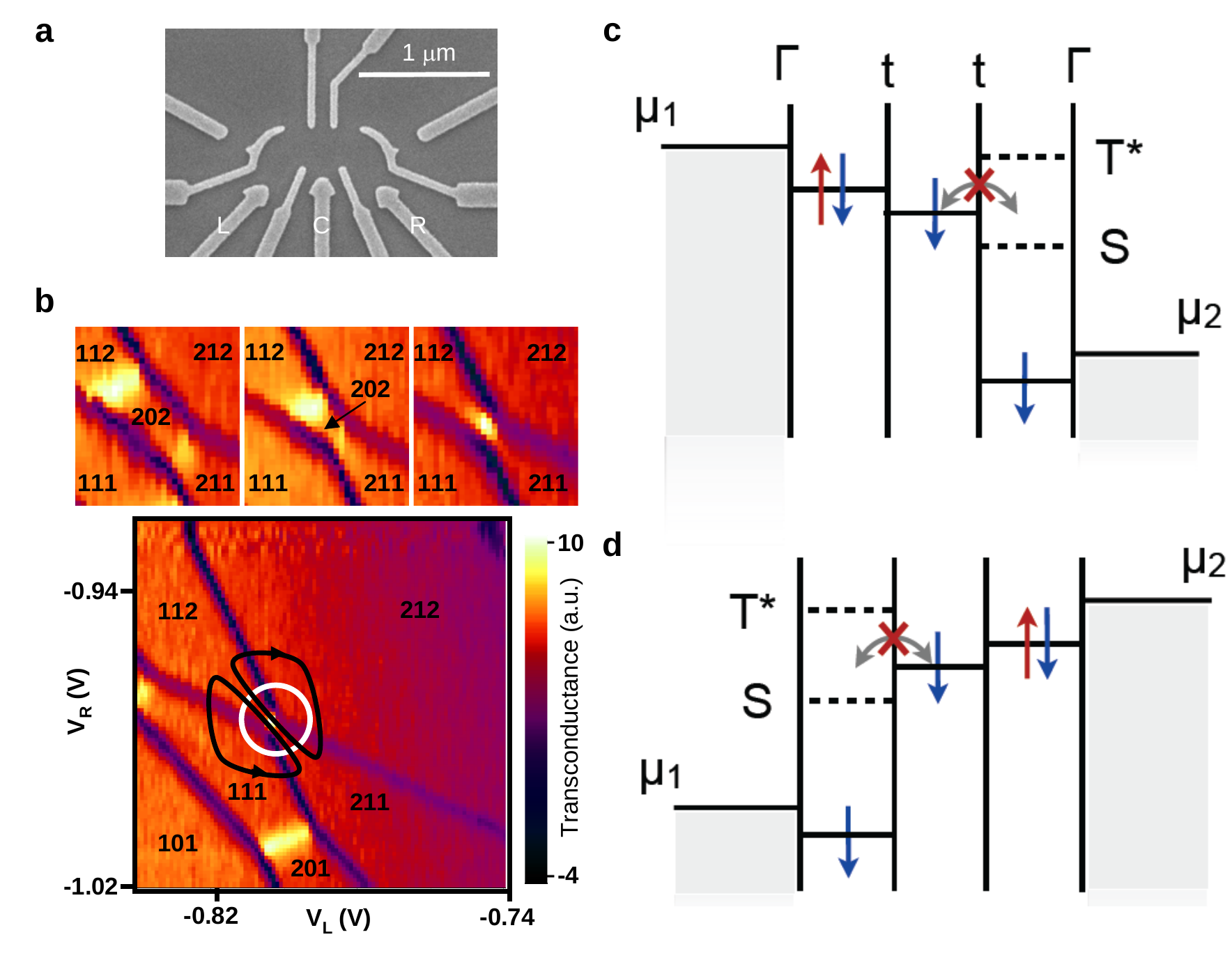}
\end{center}
\caption{
{\bf Bipolar spin blockade in a linear TQD.}
{\bf a}, Scanning electron micrograph of the sample structure. {\bf b}, (Bottom) Stability diagram of the TQD transconductance from ref~\cite{grangerPRB10} measured with the left charge detector at a fixed C gate voltage of -0.208 V while varying left (horizontal) and right (vertical) gate voltages. The tunneling sequence at quadruple points 5 and 6 of ref.~\cite{grangerPRB10} is marked by arrows. (Top row) The size of the $(2,0,2)$ region shrinks as the C gate voltage increases in equal steps from -0.214~V to -0.210~V. {\bf c}, Schematic description of spin blockade in positive bias direction. In the left and right dots, one electron is confined electrostatically so that only doubly occupied levels in left and right dots contribute to transport. The central dot can only accept one electron. If the two electrons in the centre and right dots have the same spin, current is blocked due to spin blockade. {\bf d}, Spin blockade in negative bias direction.}
\label{fig:tqdbsb_schema}
\end{figure}

The phenomenon ``spin blockade'' was first revealed in double quantum dots (DQDs)~\cite{onoSC02, Johnson2005}. For an even/odd quantum dot occupation configuration such as (0,1), current is blockaded by the Pauli exclusion principle whenever the spin entering the left dot possesses the same spin as that in the right dot~\cite{onoSC02}. Close to zero magnetic field leakage currents occur, attributed to a mixing of singlet and triplet states by the field gradient resulting from the different statistical Overhauser fields in the two dots~\cite{koppensSC05}. For fields greater than this gradient the triplet states $|{\uparrow},{\uparrow}\rangle$ and $|{\downarrow},{\downarrow}\rangle$ no longer mix with the singlet and fully restore the spin blockade. We demonstrate the novel features of spin blockade and leakage currents that occur in a TQD where the participation of coherent superpositions is revealed to be essential.
 
 
The TQD defined electrostatically in a GaAs/AlGaAs heterostructure is shown in \subfig{tqdbsb_schema}{a}. The system is tuned to the regime relevant for spin qubits bounded by electronic occupations ($N_\text{L}$,$N_\text{C}$,$N_\text{R}$)=(1,0,1) and (2,1,2), where $N_i$ is the number of electrons on the left, centre, and right quantum dots respectively, see \subfig{tqdbsb_schema}{b}. In this regime there exist six quadruple points (QPs) where four charge configurations are degenerate~\cite{grangerPRB10}. We focus on the behaviour at two QPs associated with the configurations (1,1,1), (2,1,1), (2,0,2), (1,1,2) and (2,1,2), (2,1,1), (2,0,2), (1,1,2). The experimental results are compared to theoretical calculations of the current using a master equation for the reduced density matrix of the TQD.

\begin{figure}
\begin{center}
\includegraphics[width=6.5in,clip]{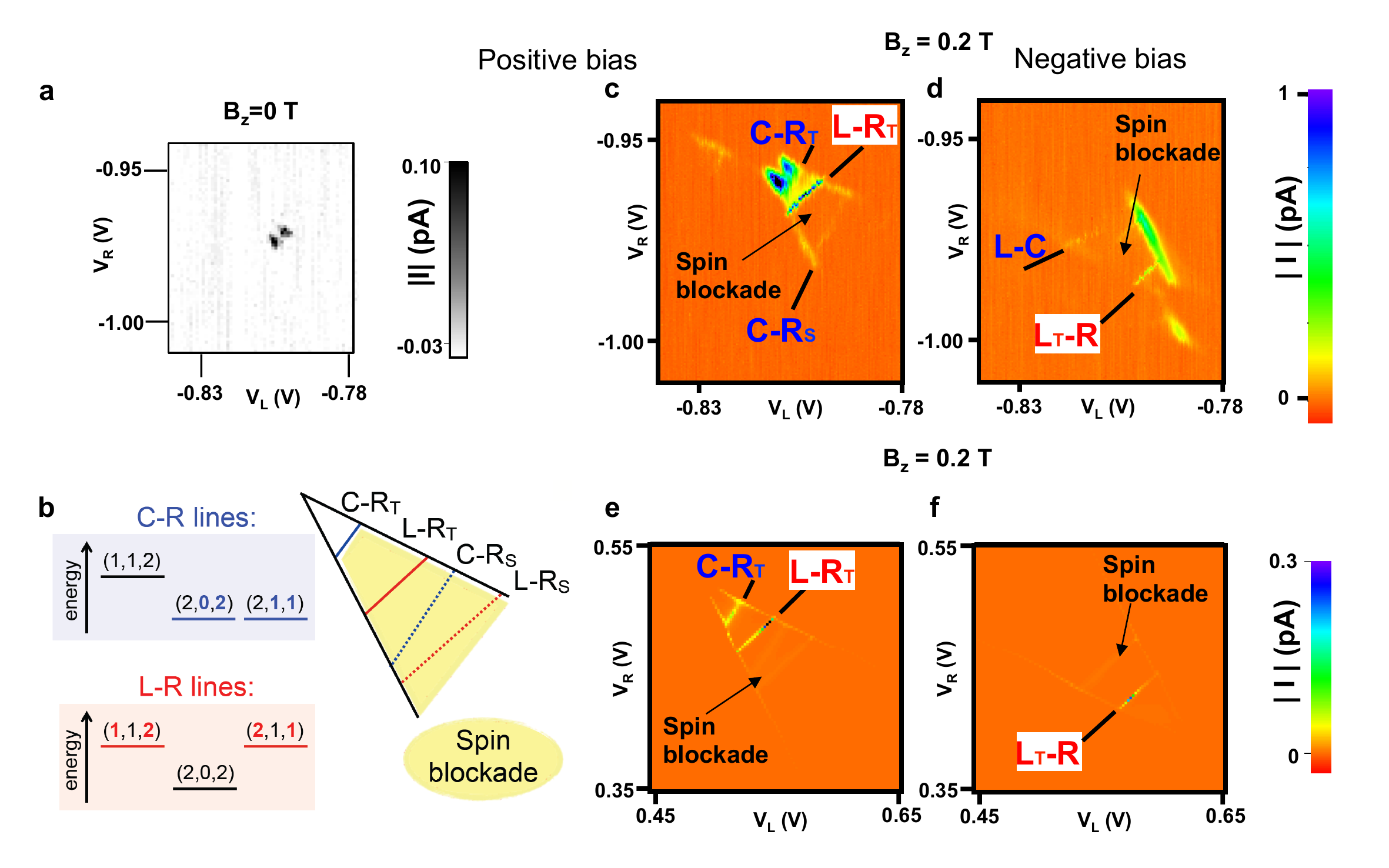}
\end{center}
\caption{
{\bf Bipolar spin blockade in the TQD.} \textbf{a,} The current  through the TQD flows only at two spots (quadruple points 5 and 6 of ref.~\cite{grangerPRB10}) when the bias is 0.1~mV at zero magnetic field. \textbf{b}. Different resonant conditions for the chemical potentials associated with the (1,1,2), (2,1,1), and (2,0,2) charge distributions are shown, and their corresponding locations are depicted schematically in the transport triangle for positive bias. \textbf{c-f} figures are at a larger bias of 0.5~mV of either polarity in the presence of a magnetic field of 0.2~T parallel to the 2DEG. For positive (c and e) and negative (d and f) bias, both the experimental (c and d) and theoretical (e and f) results are shown. In a finite magnetic field, spin-flip processes are less effective than at $B_{z}{=}0$~T and spin blockade occurs. It affects resonances involving the singlet state in the drain dot, for instance, L-R$_{\mathrm{S}}$ and C-R$_{\mathrm{S}}$ lines at positive bias.  The C-R$_\mathrm{T}$ line involves the  excited triplet level in the right dot and thus constitutes the limit beyond which spin blockade no longer occurs. Small transport regions related to other QPs are seen as a small yellow patch in the upper left corner in c and as a small green patch in the lower right corner in d.}
\label{fig:tqdbsb_fwbw_sf_exptheo}
\end{figure}

At low bias (0.1 mV), current only flows at two small spots (see Fig.~\ref{fig:tqdbsb_fwbw_sf_exptheo}a)~\cite{grangerPRB10}. If the bias is increased, the transport region expands into a triangle or quadrangle shape (see Supplementary Information, S3).  Figure~\ref{fig:tqdbsb_fwbw_sf_exptheo}c,d shows this for transport measurements made in a small magnetic field. The current is dramatically suppressed in {\it both} bias directions except along marked resonance lines. The underlying insulating behaviour can be considered an extension from the DQD spin blockade phenomenon since the TQD circuit in this regime is equivalent to two back to back DQD spin blockade rectifiers (see Fig.~\ref{fig:tqdbsb_schema}c,d), forming a TQD ``spinsulator.''

\begin{figure}
\begin{center}
\includegraphics[width=6in,clip]{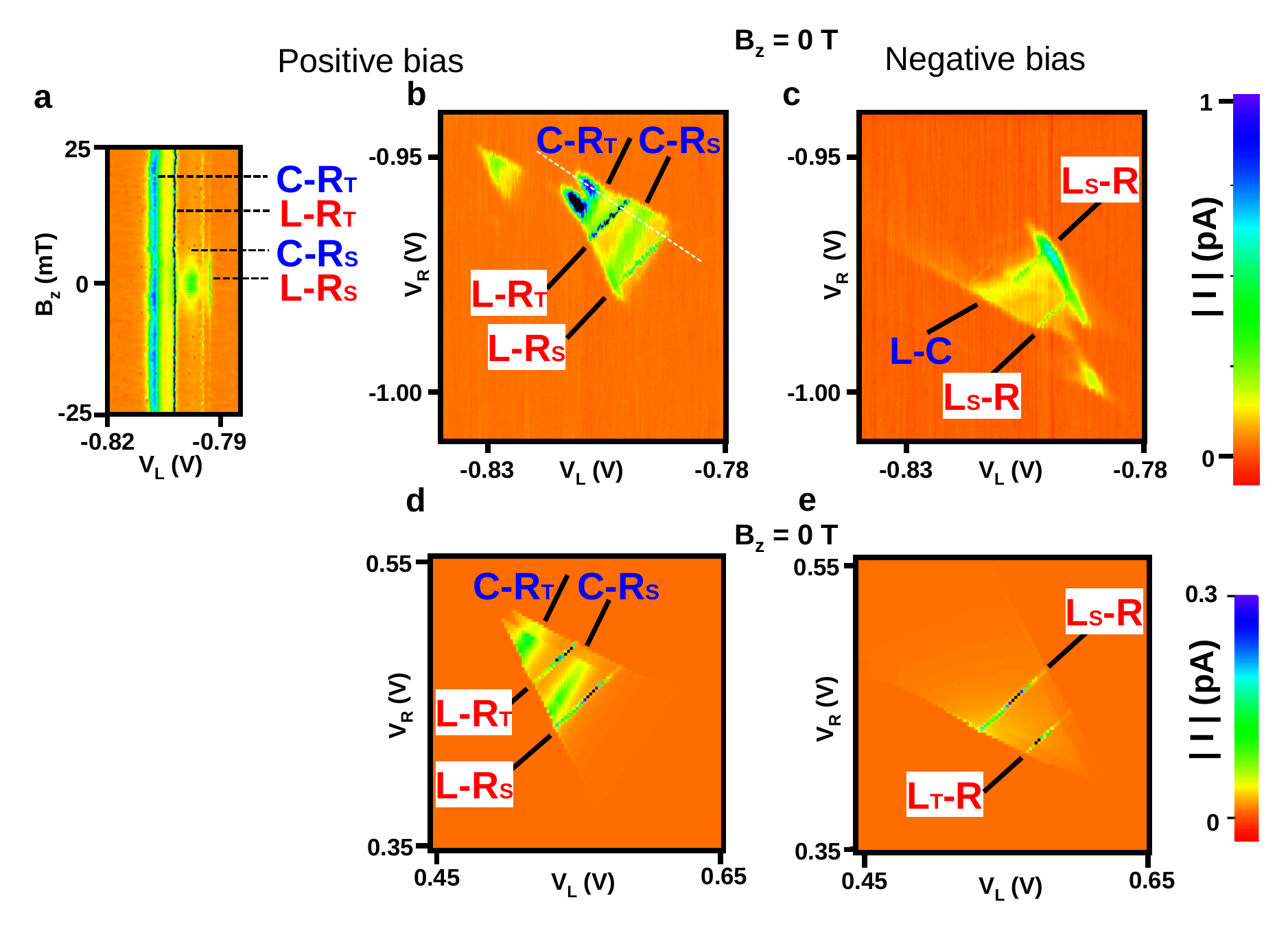}
\end{center}
\caption{
{\bf Leakage current through the TQD at quadruple points 5 and 6 of ref.~\cite{grangerPRB10} for zero magnetic field.} \textbf{a,} Magnetic field dependence of the leakage current measured with a bias of +0.5 mV by sweeping $V_\text{L}$ and $V_\text{R}$ along the white dashed line in b. The dotted lines indicate the positions of the various resonances also labeled in b. \textbf{b-e} figures are at a 0.5~mV bias of either polarity. Both for positive (b and d) and negative (c and e) bias and for the experimental (b and c) and theoretical (d and e) results, one can clearly distinguish resonance lines with two different slopes: the L-R resonance lines, where (2,1,1) and (1,1,2) states are on resonance, and the steeper C-R resonance lines that occur when the energy of states (2,1,1) and (2,0,2) are aligned. The shallow L-C lines are observed experimentally, but do not appear theoretically for this choice of parameters. Beyond the two transport triangles, there are small transport regions related to other QPs, as seen experimentally as a green triangle in the upper left corner in b and as a green patch in the lower right corner in c.} 
\label{fig:tqdbsb_fwbw_sf0_exptheo}
\end{figure}

In Fig.~\ref{fig:tqdbsb_fwbw_sf0_exptheo}b,c, we show measurements at zero magnetic field. Small leakage current contributions are observed throughout the triangular region  dominated by additional resonance lines. 
 These prominent features are accurately reproduced in the theoretical model (Fig.~\ref{fig:tqdbsb_fwbw_sf0_exptheo}d,e). A direct comparison of the slopes of the lines with the charge transfer lines in the stability diagram identifies the relevant resonant quantum dots (see Fig.~\ref{fig:tqdbsb_fwbw_sf_exptheo}b and Supplementary Information, S3).  For clarity of the underlying physics we invoke the notation of singlet and triplet states in the explanations when describing the state of two electrons in a particular dot. We stress, however, that we are dealing with interacting three and four electron states. We are able to characterize the lines in two ways. Firstly we note that a magnetic field of 5 mT is sufficient to suppress certain lines (see Fig.~\ref{fig:tqdbsb_fwbw_sf0_exptheo}a) namely L-R$_\mathrm{S}$ and C-R$_{\mathrm{S}}$ where L, R, and C refer to the left, right and centre dots and the subscripts, S or T identify whether the doubly occupied state is a singlet or triplet. This is consistent with estimates of the statistical Overhauser field gradient and by 
 analogy with the DQD scenario this field dependence identifies which lines are subject to spin blockaded events. Secondly we can characterize the lines by whether the dots at resonance involve the centre and an edge dot, i.e. C-R$_\mathrm{S(T)}$ and L-C lines, or whether only edge dots are involved, i.e. the L-R lines. C-R and L-C lines are analogous to those observed in DQDs, see Figs. 2 and 3. The L-R lines on the contrary, are specific of quantum coherent triple quantum dot systems\cite{gaudreauNAPHYS12} as they involve a resonant charge transfer between non adjacent dots. They occur when the (2,1,1) and (1,1,2) configurations are degenerate and off-resonance with respect to the  state (2,0,2). 
Note that a description in terms of sequential tunneling  through the state (2,0,2) cannot explain such resonances. They can only be understood by invoking coherent tunneling processes, for instance left-right cotunneling events via virtual transitions to the centre dot. For our relatively strong interdot tunnel coupling, higher order contributions to coherent tunneling should be considered.

\begin{figure}
\begin{center}
\includegraphics[width=5.8in,clip]{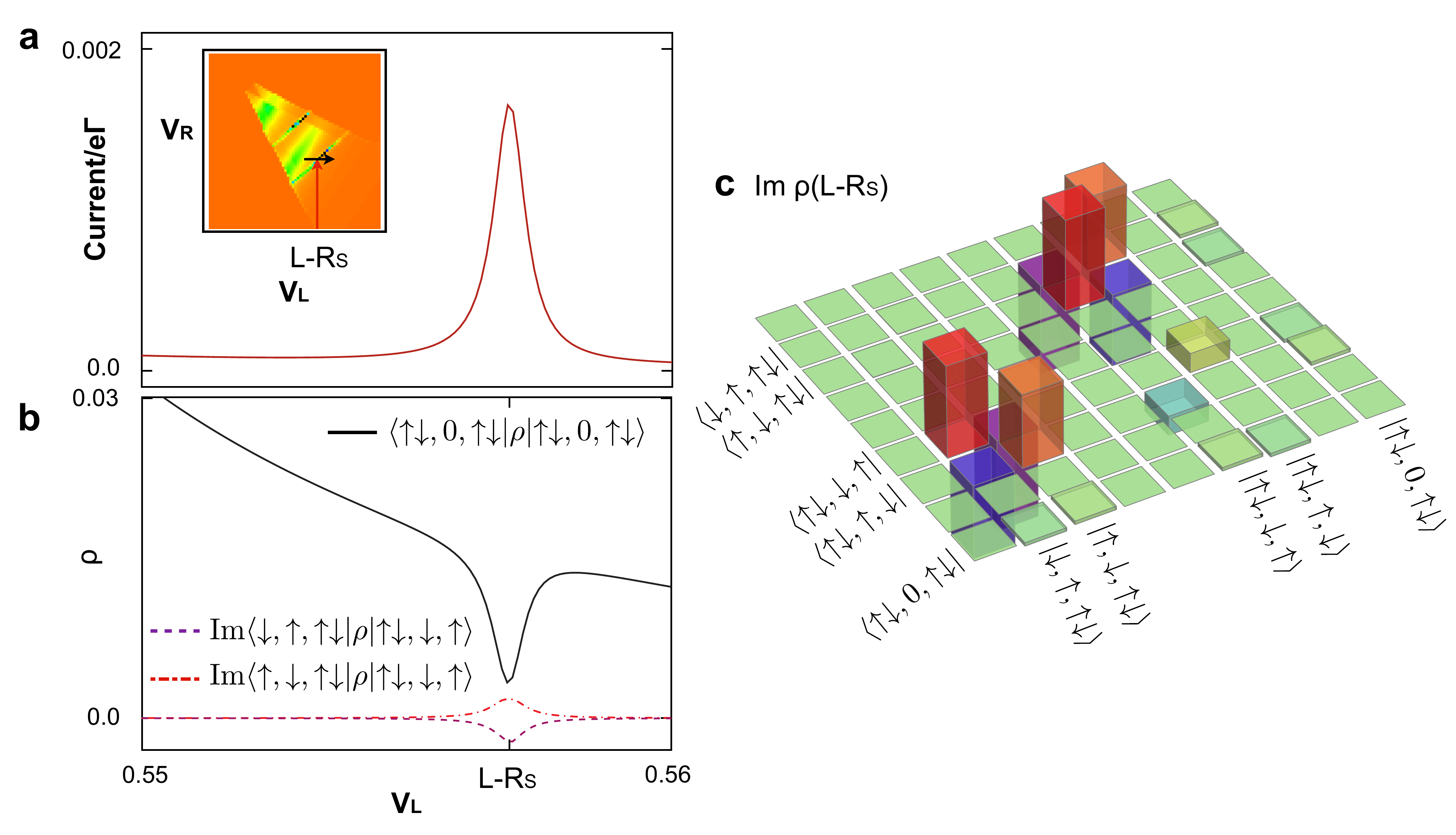}
\end{center}
\caption{
{\bf Density matrix analysis at the L-R singlet resonance line.} {\bf a}, Current $I$ (normalized by the tunneling rate $\Gamma$ to the leads) in positive bias for zero magnetic field for fixed $V_{\mathrm{R}}$ as a function of $V_{\mathrm{L}}$, indicated by  the black horizontal arrow in the inset.  {\bf b}, Selected elements of the density matrix $\rho$ along the black horizontal arrow indicated in the inset in a. The occupation of (2,0,2) -- $\langle{\uparrow}{\downarrow},0,{\uparrow}{\downarrow}|\rho|{\uparrow}{\downarrow},0,{\uparrow}{\downarrow}\rangle$ -- decreases considerably at L-R$_\text{S}$, i.e. at the L-R singlet resonance line (black solid line). Simultaneously, the red dot-dashed and purple dashed lines exemplify that the bonding between the states (2,1,1) and (1,1,2) strengthens at L-R$_\text{S}$. This bonding is represented by the imaginary part of the density matrix between the states $|{\uparrow}{\downarrow},{\downarrow},{\uparrow}\rangle$ and $|{\downarrow},{\uparrow},{\uparrow}{\downarrow}\rangle$ for the red line and between $|{\uparrow}{\downarrow},{\downarrow},{\uparrow}\rangle$ and $|{\uparrow},{\downarrow},{\uparrow}{\downarrow}\rangle$ for the purple line. The bump (dip) reflects that the {\it coherence} between these states increases in absolute value at L-R$_\text{S}$. 
{\bf c}, A steady state representation of the density matrix at the L-R$_\text{S}$ resonance. The finite off-diagonal contributions relate to the coherence between the corresponding elements. Remarkably, the prominent red, purple, blue and orange columns represent the bonding between the (2,1,1) and (1,1,2) states, which are involved in the formation of the superposition $|\Sigma\rangle$.}
\label{fig:tqdbsb_rhos}
\end{figure}

At the L-R$_\text{S}$ resonance the  $|{\uparrow}{\downarrow},0,{\uparrow}{\downarrow}\rangle$ state becomes largely depopulated, whereas the current nevertheless increases (see Fig. 4). The underlying mechanism for those L-R$_\mathrm{S}$ resonances to appear can be understood as follows: At magnetic fields smaller than the Overhauser  field gradient spin blockade between the centre and right dots (in positive bias) is removed by hyperfine-induced spin-flip processes e.g. from a state $|{\uparrow}{\downarrow},{\downarrow},{\downarrow}\rangle$ to $|{\uparrow}{\downarrow},{\uparrow},{\downarrow}\rangle$ or $|{\uparrow}{\downarrow},{\downarrow},{\uparrow}\rangle$. Superpositions of these states are formed as a consequence of the interference of multiple scattering events at the interdot barriers. One eigenstate turns out to be of particular importance for the transport along the L-R$_\mathrm{S}$ line which does not include any $|{\uparrow}{\downarrow},0,{\uparrow}{\downarrow}\rangle$ contribution at all. In the limit  of zero Overhauser field this state reads
\begin{equation}
{\mid}\Sigma\rangle {=} \frac{1}{2}\left({\mid}{\uparrow}{\downarrow},{\downarrow},{\uparrow}\rangle{-}{\mid}{\uparrow}{\downarrow},{\uparrow},{\downarrow}\rangle{-}
{\mid}{\downarrow},{\uparrow},{\uparrow}{\downarrow}\rangle{+}{\mid}{\uparrow},{\downarrow},{\uparrow}{\downarrow}\rangle\right).
\label{eq:tqdbsb_state}
\end{equation}
In the presence of finite inhomogeneous Overhauser fields coming from the hyperfine interaction, the superposition ${\mid}\Sigma\rangle$ acquires a perturbative mixing with triplet states but still without any participation of the state (2,0,2) (see Supplementary Information S2 for more details). 
That state has a finite tunneling rate to the drain contact, thereby opening the system to transport: current will flow from the source to the drain contact until spin blockade is restored again and the cycle repeats itself. 
Note that the blockade-lifting sequence does not involve the occupation of the intermediate state (2,0,2).
This explains the depopulation of the state (2,0,2).
The simultaneous increase of coherence between states (2,1,1) and (1,1,2) at the   L-R$_\text{S}$ resonance (manifested as an increase of the off diagonal density matrix
elements between the states which form the superposition, cf. Fig.~\ref{fig:tqdbsb_rhos}b and c) confirms our interpretation that transport occurs via the coherent superposition $|\Sigma\rangle$. Furthermore, the relaxation from the spin blockaded state $|{\uparrow}{\downarrow},{\downarrow},{\downarrow}\rangle$ into the $|\Sigma\rangle$ state is enhanced because at the L-R$_\text{S}$ resonance $|\Sigma\rangle$ is nearly degenerate with the spin blockaded states (see Supplementary Information S2). 

Remarkably, \eqref{eq:tqdbsb_state} is a swapped superposition of singlets with different charge distributions and with left and right dots both equally influenced by double occupation. \textit{The occupation of this superposition entails the direct transport of electrons from the  left  to the right  dot in forward bias voltage (from right to left in backward bias voltage)}. The similarity of such a superposition with those responsible for dark resonances observed in multilevel atoms or those predicted to exist in transport through quantum dots is clear~\cite{brandesPRL00,emaryEPL06,maria, amaha2012}. 

The L-R$_{\mathrm{T}}$ line, in contrast, is the resonance between the states $(2,1,1)$ and $(1,1,2)$ involving a singlet (triplet) level in the left (right) dot at positive bias. The increased leakage current is due to coherence between the three dots at the exact resonance between the left and right dots. Its appearance does not require hyperfine-induced spin-flips. The  state (2,0,2), does participate in the transport along this line even though it is off resonance. 

In conclusion, bipolar spin blockade has been observed for the first time in TQDs. Additional unexpected resonant lines in the transport diagrams between the edge quantum dots are explained via quantum coherent superpositions. 
The exact resonance between the edge dots acts as a coherent leakage current amplifier: charge is transferred directly from left to right dot, thereby circumventing the off-resonant centre site.

\section*{METHODS}
{\bf Experiment.}
The GaAs/AlGaAs heterostructure is grown by molecular beam epitaxy and has a density of 2.1$\times10^{11}$~cm$^{-2}$ and a mobility of 1.72$\times10^6$~cm$^2$/Vs. Ohmic contacts to the two-dimensional electron gas (2DEG) located 110~nm below the surface are made. TiAu gate electrodes are patterned by electron-beam lithography. They allow electrostatic control of the triple quantum dot (TQD). On the left and right of the TQD are two gates defining quantum point contacts (QPCs) used as charge detectors. 

Charge detection measurements are made by measuring the conductance of one of the charge detectors with a lock-in technique using a typical root-mean-square modulation in the 0.05-0.1~mV range. The QPC detector conductance is tuned to below 0.1~e$^2$/h. Current measurements are made by applying a bias up to 1.5~mV of either polarity across the TQD and measuring the resulting direct current with a current preamplifier in the voltage plane defined by the left and right electrodes. The device is bias-cooled in a dilution refrigerator with 0.25~V on all gates. Once cold, suitable gate voltages are applied to the gates to form the TQD potential.  The electron temperature is approximately 110~mK in this system.

{\bf Theory.}
We model the TQD, the leads and the coupling between them with an Anderson-like Hamiltonian which includes 
the coherent tunneling $t_{ij}$ between the dots, the static magnetic field $B_{\text{z}}$, the coupling to the leads by a rate $\Gamma$ and the leads themselves.
In order to obtain the current through the TQD we resort to standard techniques using a master equation for the reduced density matrix, see e.g. ref.~\cite{blum} and Supplementary Information S1. Our coherent interdot tunneling calculations include virtual transitions through intermediate states to infinite order in perturbation theory. As a result coherent cotunneling (second order in perturbation theory) is included within our model. Cotunneling processes to the leads cannot be responsible for L-R resonance features, therefore only sequential tunneling processes  through the contact barriers were included.

As we are interested in the stationary current through the TQD, we solve the set of equations of the reduced density matrix algebraically and then calculate the current (see Supplementary Information S1 for more details).
For an approximate modeling of the spin-flip processes induced by hyperfine interaction we included a {\it phenomenological} spin-flip rate into the equations, and we included a finite inhomogeneous Overhauser field into the Hamiltonian for the magnetic field. Although not providing a microscopically rigorous description, the spin-flip rates qualitatively reproduce very well the effects expected due to hyperfine coupling in quantum dots. We set the spin relaxation rate $T_{1}\approx1~\mu$s and the spin decoherence time $T_{2}^{*}\approx10$ ns. Interdot tunneling time is of the order of 0.1~ns. These parameters are consistent with those provided by the experimental evidence 
\cite{gaudreauNAPHYS12}. A detailed description of the parameters considered is given in the Supplementary Information S1.




\textbf{Acknowledgements}\\
We thank P.~Hawrylak and C.~Y.~Hsieh for discussions. M.~B., R.~S. and G.~P. acknowledge financial support from the Spanish Ministry of Education through Grant No. MAT2011-24331 (MEC) and from the Marie Curie Initial Training Network under Grant No. 234970 (EU). M. B. and R. S. were supported by the Consejo Superior de Investigaciones Cient\'ificas through the JAE and JAE-Doc programs,  cofinanced by the Fondo Social Europeo. G.~G.~acknowledges funding from the National Research Council Canada -- Centre national de la recherche scientifique collaboration and Canadian Institute for Advanced Research.  A.~S.~S.~acknowledges funding from Natural Sciences and Engineering Research Council of Canada and Canadian Institute for Advanced Research. \\

\newpage
\textbf{Author contributions}\\
M.B. and R.S. performed the theoretical calculations. A.K. fabricated the TQD device. Z.R.W. optimized and grew the 2DEG heterostructure.  G.G., L.G., and S.A.S. performed the experiments and analysis. M.P.L. assisted with these measurements and analysis. P.Z. assisted with the experiments.  M.B., G.G., R.S., A.S.S. and G.P. wrote the paper.  A.S.S. and G.P. supervised the experimental and theoretical components of the collaboration.\\

\textbf{Author information}\\
The authors declare no competing financial interests. 


\newpage
\begin{center}
\section*{\underline{SUPPLEMENTARY INFORMATION}}
\end{center}

\section*{S1. Theoretical model}
We model the TQD system and the leads by an Anderson-like Hamiltonian that reads
$\mathcal{H} = \mathcal{H}_{\text{TQD}}+\mathcal{H}_{t_{ij}}+\mathcal{H}_{\text{B}}+\mathcal{H}_{\text{TQD-leads}}+\mathcal{H}_{\text{leads}}$,
where the individual terms are
\begin{align}
\label{ham}
\mathcal{H}_{\text{TQD}} &= \sum_{ik\sigma}\epsilon_{ik\sigma}\hat{c}^{\dagger}_{ik\sigma}
\hat{c}_{ik\sigma}+\sum_{i}U_{i}\hat{n}_{i\uparrow}\hat{n}_{i\downarrow}
+\frac{1}{2}\sum_{i\neq j}V_{ij}\hat{n}_{i}\hat{n}_{j}+\sum_{i}J_i{\bf S}_{i0}{\bf S}_{i1}\nonumber\\
\mathcal{H}_{t_{ij}} &= -\sum_{i\neq j,k,\sigma}t_{ij}(\hat{c}^{\dagger}_{ik\sigma}
\hat{c}_{jk\sigma}+\hat{c}^{\dagger}_{jk\sigma}\hat{c}_{ik\sigma})\nonumber\\
\mathcal{H}_{\text{B}} &= \sum_{i}g\mu_{\mathrm{B}}(B_z+B_{i,\text{nucl}})S_{z,ik}=\sum_{i,k}\Delta_{i}S_{z,ik}\\
\mathcal{H}_{\text{TQD-leads}} &= \sum_{l\in{\rm L,R},qk\sigma}\gamma_{l}(\hat{d}^{\dagger}_{lq\sigma}
\hat{c}_{lk\sigma}+\hat{c}^{\dagger}_{lk\sigma}\hat{d}_{lq\sigma})\nonumber\\
\mathcal{H}_{\text{leads}} &= \sum_{l\in{\rm L,R},q\sigma}\varepsilon_{lq}\hat{d}^{\dagger}_{lq\sigma}
\hat{d}_{lq\sigma}.\nonumber
\end{align}
The first term represents the TQD itself, with $\epsilon_{ik\sigma}$ being the energy of an electron with spin $\sigma$ occupying the ground ($k{=}0$) or excited ($k{=}1$) state in dot $i=1,2,3$. The energy separation between excited and ground levels is given by $\varepsilon{=}\epsilon_{i1\sigma}{-}\epsilon_{i0\sigma}$. The excited level of the centre dot is not considered. $U_{i}$ is the on-site Coulomb interaction energy, $V_{ij}$ are the interdot Coulomb interaction energies; we set $V_{12}{=} V_{23}{=}V{\neq} V_{13}$. Intradot exchange interaction is given by $J_i$, and the spin operators are $\mathbf{S}_{ik} {=} \frac{1}{2}\sum_{\sigma\sigma'}c_{ik\sigma}^{\dagger}\sigma_{\sigma\sigma'}c_{ik\sigma'}$ with $\sigma_{\sigma\sigma'}$ being the Pauli spin matrices.  The second term describes the coherent tunneling between the dots, where $t_{12,23}=t$ and $t_{13}=0$, so {\it no direct tunneling is possible from dot 1 to dot 3}. The effect of a static magnetic field $B_{z}$ i
 s described in the third term, where we include the z-component of the Overhauser field induced by the nuclei of the host material, $B_{i,\text{nucl}}$; $g$ is the electron $g$-factor and $\mu_{\mathrm{B}}$ the Bohr magneton. The fourth term describes the tunneling between dot 1 and the left lead and between dot 3 and the right lead with an amplitude $\gamma_{l}$, and finally the last term describes the leads themselves, where $\varepsilon_{lq}$ is the energy of an electron in lead $l$. 
The creation and annihilation operators for an electron on dot $i$ with spin $\sigma$ are given by $c_{ik\sigma}^{\dagger}, c_{ik\sigma}$, and for an electron in lead $l$ by $d^{\dagger}_{lq\sigma},d_{lq\sigma}$. 

In the experiment, the current is measured for a fixed centre gate voltage while varying the left and right gate voltages. A change in the left gate voltage however does not only affect the energy levels in the left dot, but due to cross capacitances it also --- albeit more weakly --- affects the neighbouring dots. The energy levels of the dots can be written as linear functions of the affecting gate voltages. Following therefore the scheme in ref.~\cite{gaudreauPRL06} and considering cross capacitances as proposed by the experiment, we write the energy levels $\epsilon_{i}$ of the dots $i{=}1,2,3$ as
$\epsilon_{i}{=}\mathcal{C}_i - \alpha_i V_{\text{L}} - \beta_i V_{\text{R}},$
where $\mathcal{C}_{1,2,3}$ are constants that provide an overall energy shift, and the conversion parameters $\alpha_i$ and $\beta_i$ are written in eV/V.

The current is measured around the quadruple points (QPs) 5 and 6, see ref.~\cite{grangerPRB10}. At these points the following states with the specified electron numbers in the left, centre and right dots, $(N_\text{L},N_\text{C},N_\text{R})$, are resonant: (1,1,1), (2,1,1), (2,0,2), (1,1,2) at QP 5 and (1,1,2), (2,1,2), (2,1,1), (2,0,2) at QP 6. 
For the doubly occupied levels in the left and right dots we make the following assumptions:
in positive bias (negative bias), the left (right) dot -- i.e. the dot connected to the electron source -- accepts an additional incoming electron, so that the two electrons occupy a {\it singlet} state $|{\uparrow}{\downarrow}\rangle$. The higher levels, such as excited triplets $|{\uparrow}{\uparrow}^{*}\rangle$ , $|{\downarrow}{\downarrow}^{*}\rangle$ and $1/\sqrt{2}(|{\uparrow}{\downarrow}^{*}\rangle {+}|{\downarrow}{\uparrow}^{*}\rangle)$ or excited singlet 
$1/\sqrt{2}(|{\uparrow}{\downarrow}^{*}\rangle{-}|{\downarrow}{\uparrow}^{*}\rangle)$ states are not accessible for an incoming electron. In contrast, the right (left) dot in positive (negative) bias is modeled in such a way that not only the singlet state $|{\uparrow}{\downarrow}\rangle$, but also the energetically higher excited states $|{\uparrow}{\uparrow}^{*}\rangle$, $|{\downarrow}{\downarrow}^{*}\rangle$, $1/\sqrt{2}(|{\uparrow}{\downarrow}^{*}\rangle{+}|{\downarrow}{\uparrow}^{*}\rangle)$  and
$1/\sqrt{2}(|{\uparrow}{\downarrow}^{*}\rangle{-}|{\downarrow}{\uparrow}^{*}\rangle)$ can participate in transport.
Under these assumptions the full basis of states for the present problem contains 58 states. 
In the positive bias direction these states are
\begin{align}
|1,1,1\rangle &= |\sigma,\sigma',\sigma''\rangle & |1,1,1^*\rangle &= |\sigma'',\sigma',\sigma^*\rangle\nonumber\\ 
|2,1,1\rangle &= |S,\sigma',\sigma\rangle & |2,1,1^*\rangle &= |S,\sigma',\sigma^*\rangle\nonumber\\
|2,0,2\rangle &= |S,0,S\rangle & |2,0,2^*\rangle &= |S,0,T^*(S^*)\rangle\nonumber\\
|1,1,2\rangle &= |\sigma,\sigma',S\rangle & |1,1,2^*\rangle &= |\sigma,\sigma',T^*(S^*)\rangle\nonumber\\
|2,1,2\rangle &= |S,\sigma,S\rangle & |2,1,2^*\rangle &= |S,\sigma,T^*(S^*)\rangle,
\end{align}
with $\{\sigma,\sigma',\sigma''\}{=}\{{\uparrow},{\downarrow}\}$ and $\sigma^*$ being an electron in an excited level. $S{=}|{\uparrow}{\downarrow}\rangle$ denotes the doubly occupied singlet level, $T^*{=}|{\uparrow}{\uparrow}^*\rangle, |{\downarrow}{\downarrow}^*\rangle, 1/\sqrt{2}(|{\uparrow}{\downarrow}^*\rangle{+}|{\downarrow}{\uparrow}^*\rangle)$ the excited triplet levels in the right dot, and finally $|S^*\rangle{=}1/\sqrt{2}(|{\uparrow}{\downarrow}^*\rangle{-}|{\downarrow}{\uparrow}^*\rangle)$ stands for the excited singlet level in the right dot. We assume that at zero magnetic field the excited triplet and singlet levels are separated from each other by the negative exchange interaction $J$, so that the singlet level $|S^*\rangle$ is higher in energy than the triplets $|T^*\rangle$. 

In order to calculate the current we make use of the density matrix formalism, see e.g.~[\onlinecite{blum}].
For each of the basis state elements, the equation of motion for the
reduced density matrix element $\rho_{mn}$ reads, within the Born-Markov-approximation,
\begin{equation}
\dot{\rho}_{mn}(t)=-i\langle m{\mid}[\mathcal{H}_{\text{TQD}}{+}\mathcal{H}_{t_{ij}}{+}\mathcal{H}_{\text{B}},
\rho]{\mid}n\rangle
+\sum_{k\neq{n}}(\Gamma_{nk}\rho_{kk}{-}\Gamma_{kn}\rho_{nn})\delta_{mn}-\Lambda_{mn}\rho_{mn}(1{-}\delta_{mn}).
\label{eq:tqdbsb_dm}
\end{equation}
The commutator accounts for the coherent dynamics in the quantum dot array.
Transition rates $\Gamma_{mn}$ from state $|n\rangle$ to
state $|m\rangle$ are of two kinds: those due to sequential tunneling through the leads, and those due to spin-flip processes.  These events induce decoherence which is taken into account
in the term $\Lambda_{mn}{=}\frac{1}{2}\sum_{k}(\Gamma_{km}{+}\Gamma_{kn})$.
Tunneling transition rates are calculated using Fermi's golden rule
\begin{equation}
\Gamma_{mn} = \sum_{l={\rm L,R}}\Gamma_{l}\{f(E_{m}{-}E_{n}{-}\mu_{l})\delta_{N_m,N_{n}{+}1}+[1{-}f(E_{n}{-}E_{m}{-}\mu_{l})]\delta_{N_{m},N_{n}-1}\},
\label{eq:tqdbsb_rates}
\end{equation}
where $E_{m}-E_{n}$ is the energy difference between states $|m\rangle$ and $|n\rangle$ of the isolated quantum dot array, $\mu_{\rm L,R}$ are the chemical potentials of the left (1) and right (2) leads, and $\Gamma_{\rm L,R} {=}  2\pi\mathcal{D}_{\rm L,R}|\gamma_{\rm L,R}|^2$ are the tunneling rates for each lead. 
The density of states $\mathcal{D}_{\rm L,R}$ and the tunneling couplings $\gamma_{\rm L,R}$ are assumed to be energy independent. We set $\Gamma_{\rm L} {=} \Gamma_{\rm R} {=} \Gamma$. 

We calculate the stationary current through the TQD by solving the set of equations of the reduced density matrix algebraically. The current in positive bias direction (i.e. from left to right) is then given by
\begin{equation}
I_{\mathrm{R}} = e\sum_{mn}\left(\Gamma^{+}_{mn}\rho_{nn} - \Gamma^{-}_{mn}\rho_{nn}\right),
\end{equation}
where $\Gamma^{+}_{mn}$ expresses the rate of tunneling from the TQD to one lead between state $|n\rangle$ (of the type (2,1,2) or (1,1,2), corresponding to QPs 5 and 6) and state $|m\rangle$ (of the type (2,1,1) or (1,1,1)), and $\Gamma^{-}_{nm}$ analogously expresses the tunneling rate from one lead to the TQD. 

Finally, in order to approximately take into account the spin-flip processes due to hyperfine interaction, we include a phenomenological spin-flip rate into the master equation. 
The spin relaxation time $T_1$ is given by $T_1{=}(W_{\uparrow\downarrow} {+} W_{\downarrow\uparrow})^{-1}$, 
where $W_{\uparrow\downarrow}$ and $W_{\downarrow\uparrow}$
are spin-flip rates that fulfill a detailed balance condition
$W_{\downarrow\uparrow}^i {=} \exp\left(\frac{-\Delta_{i}}{k_B T}\right)W_{\uparrow\downarrow}^i$.
Here $\Delta_{i}$ is the effective Zeeman splitting in quantum dot $i$ as defined in Eq.~\eqref{ham}, $k_{\mathrm{B}}$ is the Boltzmann constant and $T$  the temperature. We assume a finite temperature of $k_{\mathrm{B}}T\approx0.0086$ meV.
$T_2$ is the spin decoherence time --- i.e., the time over which a superposition
of opposite spin states of a single electron remains
coherent. This time can be affected by spin relaxation and by the spin dephasing time $T^*_2$, i.e.
the spin decoherence time for an ensemble of spins.
Hyperfine interaction induced spin-flip of a single spin is less effective in the presence of a magnetic field. For this reason, we consider $T_1\approx 1$~$\mu$s and $T_2\approx 10$~ns for $B_z=0$, and an order of magnitude larger when a finite magnetic field is applied.
At zero external magnetic field, we set the inhomogeneous Overhauser splittings $\Delta_1 {=} 0.15\cdot10^{-3}$, $\Delta_2{=} 0.2\cdot10^{-3}$, $\Delta_3 {=} 0.1\cdot10^{-3}$ (in meV). The rest of the parameters used for the calculations (Figs.~2,~3 and ~4 in the main article) are (in meV):  $t_{12,23} {=} 0.005$, $U_1 = 3.1$, $U_3 {=} 2.2$, $V {=} 0.5$, $V_{13} {=} 0.25$, $\varepsilon {=} 0.6$, $J {=}{-}0.02$, $\Gamma_{\mathrm{L,R}}{ }= 0.001$ (${\approx}1.5$ GHz), $k_{\mathrm{B}}T{=}0.0086$. 

\section*{S2. Eigenvalues at the singlet L-R resonance}
\label{sec:eig}
\begin{figure}
\begin{center}
\includegraphics[width=5.0in,clip]{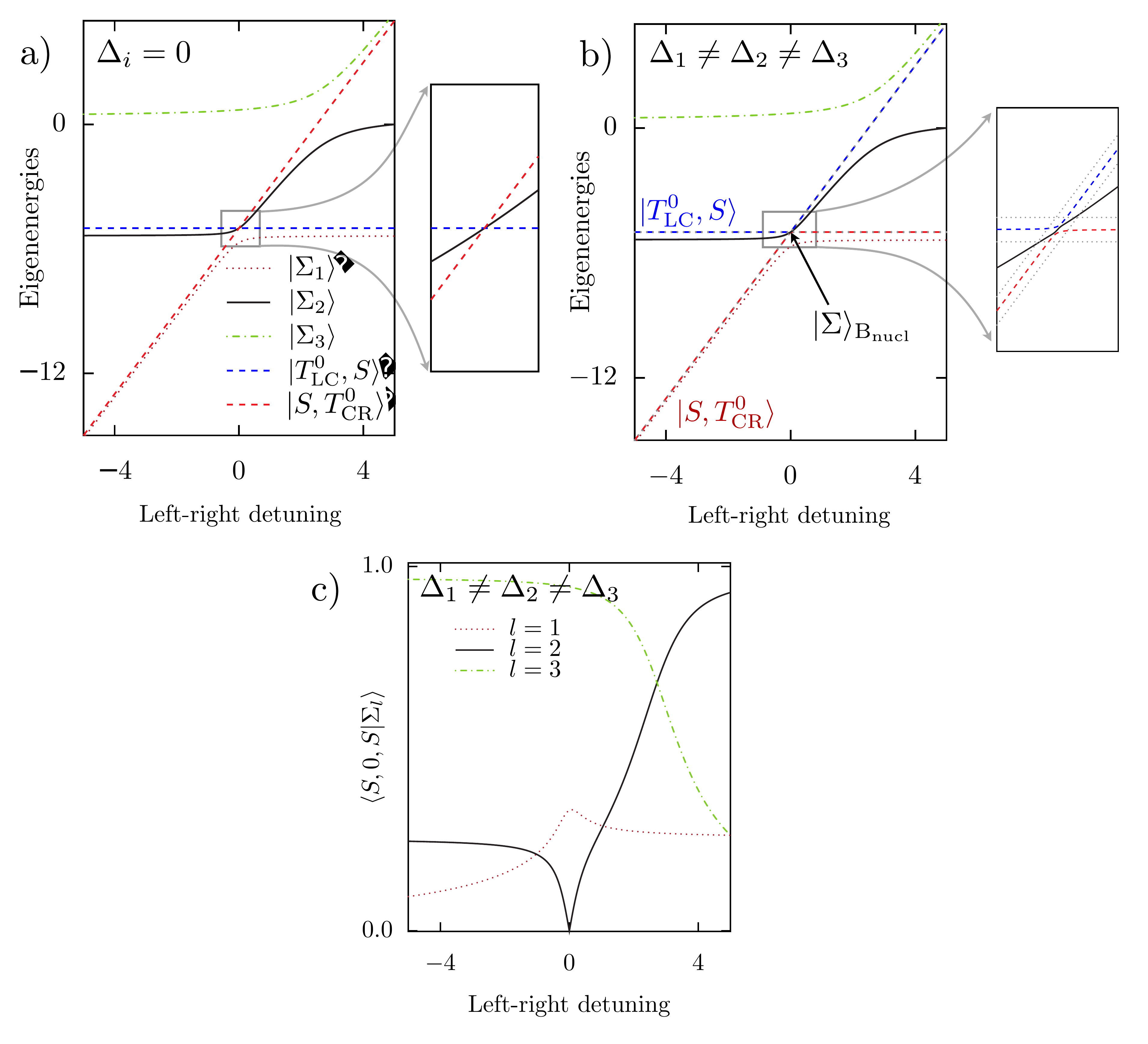}
\end{center}
\caption{
{\bf Eigenvalues around the singlet L-R resonance at zero external magnetic field.} {\bf a,} Eigenenergies of the closed system as a function of the L-R detuning for zero Zeeman splittings. The right and centre dots are not in resonance. 
The eigenenergies of the additional triplet states coincide with those of $|T_{\mathrm{LC}}^{0},S\rangle$ and $|S,T_{\mathrm{CR}}^{0}\rangle$, as defined in the text. Note in the zoomed region the crossing between the eigenstate $|\Sigma_2\rangle$ (black solid), $|S,T_{\mathrm{CR}}^0\rangle$ and $|T_{\mathrm{LC}}^0,S\rangle$ (dashed red and blue lines). 
{\bf b,} Eigenenergies of the closed system for Overhauser-induced inhomogeneous Zeeman splittings. In the zoomed region on the right hand side one can appreciate the anticrossing in the energies due to the nuclear-induced inhomogeneous splittings. At zero L-R detuning, the eigenstate $|\Sigma_2\rangle$ (Eq.~\eqref{eq:sigma2}) mixes with the triplet states $|T_{\mathrm{LC}}^{0},S\rangle$ and $|S,T_{\mathrm{CR}}^{0}\rangle$ to form $|\Sigma\rangle_{B_{\rm{nucl}}}$, but, as in the homogeneous case, $|\Sigma_2\rangle$ does not mix with the state $|S,0,S\rangle$. The eigenenergies of states $|S,T_{\mathrm{CR}}^\pm\rangle$ and $|T_{\mathrm{LC}}^\pm,S\rangle$ are represented by the unmixed dotted grey lines.
{\bf c,} Contribution of the state $|S,0,S\rangle$ to eigenstates $|\Sigma_l\rangle$ for inhomogeneous Overhauser splittings. At zero L-R detuning the state $|\Sigma\rangle_{B_{\rm{nucl}}}$ only contains contributions of states (2,1,1) and (1,1,2) up to first order in the inhomogeneity of the splittings $\Delta_{i}$.}
\label{fig:eigval202}
\end{figure}

In order to understand the drop of occupation of the state $|S,0,S\rangle$ at the {\it singlet} L-R resonance, see Fig. 4 of the main text, we analyze the eigenstates of the closed system in the absence of a magnetic field. We consider the states that contribute to transport: $|S,{\sigma},{\sigma'}\rangle$, $|S,0,S\rangle$, $|{\sigma},{\sigma'},S\rangle$, with $\{\sigma,\sigma'\}{=}\{{\uparrow},{\downarrow}\}$ and $S{=}|{\uparrow}{\downarrow}\rangle$. Let us define important states (without normalizing them, for the sake of simplicity):
\begin{align*}
|S_{\mathrm{LC}},S\rangle&=|{\uparrow},{\downarrow},{\uparrow}{\downarrow}\rangle-|{\downarrow},{\uparrow},{\uparrow}{\downarrow}\rangle && |T_{\mathrm{LC}}^0,S\rangle=|{\uparrow},{\downarrow},{\uparrow}{\downarrow}\rangle+|{\downarrow},{\uparrow},{\uparrow}{\downarrow}\rangle && |S,T_{\mathrm{CR}}^0\rangle=|{\uparrow}{\downarrow},{\uparrow},{\downarrow}\rangle+|{\uparrow}{\downarrow},{\downarrow},{\uparrow}\rangle\\
|S,S_{\mathrm{CR}}\rangle&=|{\uparrow}{\downarrow},{\uparrow},{\downarrow}\rangle-|{\uparrow}{\downarrow},{\downarrow},{\uparrow}\rangle && |T_{\mathrm{LC}}^+,S\rangle=|{\uparrow},{\uparrow},{\uparrow}{\downarrow}\rangle && |S,T_{\mathrm{CR}}^+\rangle=|{\uparrow}{\downarrow},{\uparrow},{\uparrow}\rangle\\
|S,0,S\rangle&=|{\uparrow}{\downarrow},0,{\uparrow}{\downarrow}\rangle && |T_{\mathrm{LC}}^-,S\rangle=|{\downarrow},{\downarrow},{\uparrow}{\downarrow}\rangle && |S,T_{\mathrm{CR}}^-\rangle=|{\uparrow}{\downarrow},{\downarrow},{\downarrow}\rangle
\end{align*}
The notation $S_{ij}$, $T_{ij}^\alpha$ refers to singlet and triplet superpositions formed by electrons in different quantum dots, respectively.

Let us first neglect the contribution of the inhomogeneous Overhauser field, which will be considered perturbatively later.
The eigenstates are then $|T_{\mathrm{LC}}^\alpha,S\rangle$,  $|S,T_{\mathrm{CR}}^\alpha\rangle$ and three linear combinations of the three singlets that we denote as $|\Sigma_l\rangle$. All $|\Sigma_l\rangle$ contain a contribution of $|S,0,S\rangle$ which depends on the detuning and the interdot hopping: 
\begin{eqnarray}
|\Sigma_1\rangle&=&\gamma_1|S,S_{\mathrm{CR}}\rangle+\eta_1|S_{\mathrm{LC}},S\rangle+\delta_1|S,0,S\rangle\label{eq:sigma1}\\
|\Sigma_2\rangle&=&\gamma_2|S,S_{\mathrm{CR}}\rangle+\eta_2|S_{\mathrm{LC}},S\rangle+\delta_2|S,0,S\rangle\label{eq:sigma2}\\
|\Sigma_3\rangle&=&\gamma_3|S,S_{\mathrm{CR}}\rangle+\eta_3|S_{\mathrm{LC}},S\rangle+\delta_3|S,0,S\rangle\label{eq:sigma3}
\end{eqnarray}
Note that spin blockade avoids the overlap of states $|T_{\mathrm{LC}}^\alpha,S\rangle$,  $|S,T_{\mathrm{CR}}^\alpha\rangle$.

Of special importance is $|\Sigma_2\rangle$ for two reasons: at the L-R resonance condition
\begin{itemize}
\item[({$i$})] it crosses the triplet states, cf.~Fig.~\ref{fig:eigval202}(a),
\item[({$ii$})]  at this resonance, $\gamma_2=-\eta_2$ and $\delta_2{=}0$, i.e. the contribution of $|S,0,S\rangle$ to the superposition vanishes. 
\end{itemize}
Then, ${\mid}\Sigma_2\rangle={\mid}\Sigma\rangle$ as defined in the main text:
\begin{equation}
{\mid}\Sigma\rangle {=} \frac{1}{2}\left({\mid}S_{\mathrm{LC}},S\rangle{-}{\mid}S,S_{\mathrm{CR}}\rangle\right).
\label{eq:tqdbsb_state_sup}
\end{equation}

In the presence of hyperfine interaction, the nuclei induce Overhauser fields which are slightly different within the different quantum dots. These Overhauser fields give rise to inhomogeneous effective Zeeman splittings $\Delta_i$ in the TQD. Those mix the singlet and triplets with the exception of those with parallel spins, $|T_{\mathrm{LC}}^\pm,S\rangle$ and  $|S,T_{\mathrm{CR}}^\pm\rangle$. 
As compared to the homogeneous case, instead of crossings, we now obtain anticrossings in the energy spectrum around the resonance condition, see Fig.~\ref{fig:eigval202}(b). At these anticrossings, the former states $|\Sigma_2\rangle$ and the triplet states $|T_{\mathrm{LC}}^0,S\rangle$ and $|S,T_{\mathrm{CR}}^0\rangle$ mix and are not eigenstates of the Hamiltonian anymore. Importantly for our discussion, to leading order in a perturbative expansion, the superposition $|\Sigma\rangle$ responsible for the spin blockade removal does not mix with $|S,0,S\rangle$. Concretely, it reads:
\begin{equation}
{\mid}\Sigma\rangle_{B_{\rm{nucl}}}={\mid}\Sigma\rangle+(\Delta_1{-}\Delta_2){\mid}S,T_{\mathrm{CR}}^0\rangle+(\Delta_2{-}\Delta_3){\mid}T_{\mathrm{LC}}^0,S\rangle+{\cal O}(B_{\text{nucl}}^2). 
\label{eq:sigma2a}
\end{equation}
Close to resonance, this state crosses the states ${\mid}S,T_{\mathrm{CR}}^\pm\rangle$ that are responsible for spin blockade, as shown in Fig.~\ref{fig:eigval202}. 

Our analysis suggests hence that the {\it lifting} of spin blockade occurs via the spin-flip decay of the blocking states, ${\mid}S,T_{\mathrm{CR}}^\pm\rangle$ into 
$|\Sigma\rangle_{B_{\rm{nucl}}}$.
 The latter has a finite tunneling rate to the drain contact, therefore opening the system to transport: current will flow from the source to the drain contact until spin blockade is restored again. Note that the blockade lifting transition does not involve the occupation of the state $|S,0,S\rangle$. The sharp dip in $\langle S,0,S|\Sigma\rangle_{B_{\rm{nucl}}}$ (cf. Fig.~\ref{fig:eigval202}(c)) is therefore consistent with the occupation of $|S,0,S\rangle$ in the stationary solution of the transport configuration, cf. Fig. 4 of the main article. There, the minimum remains finite due to the contribution of the other current channels in which $|S,0,S\rangle$ participates once spin blockade is removed.

\begin{figure}
\begin{center}
\includegraphics[width=3.8in,clip]{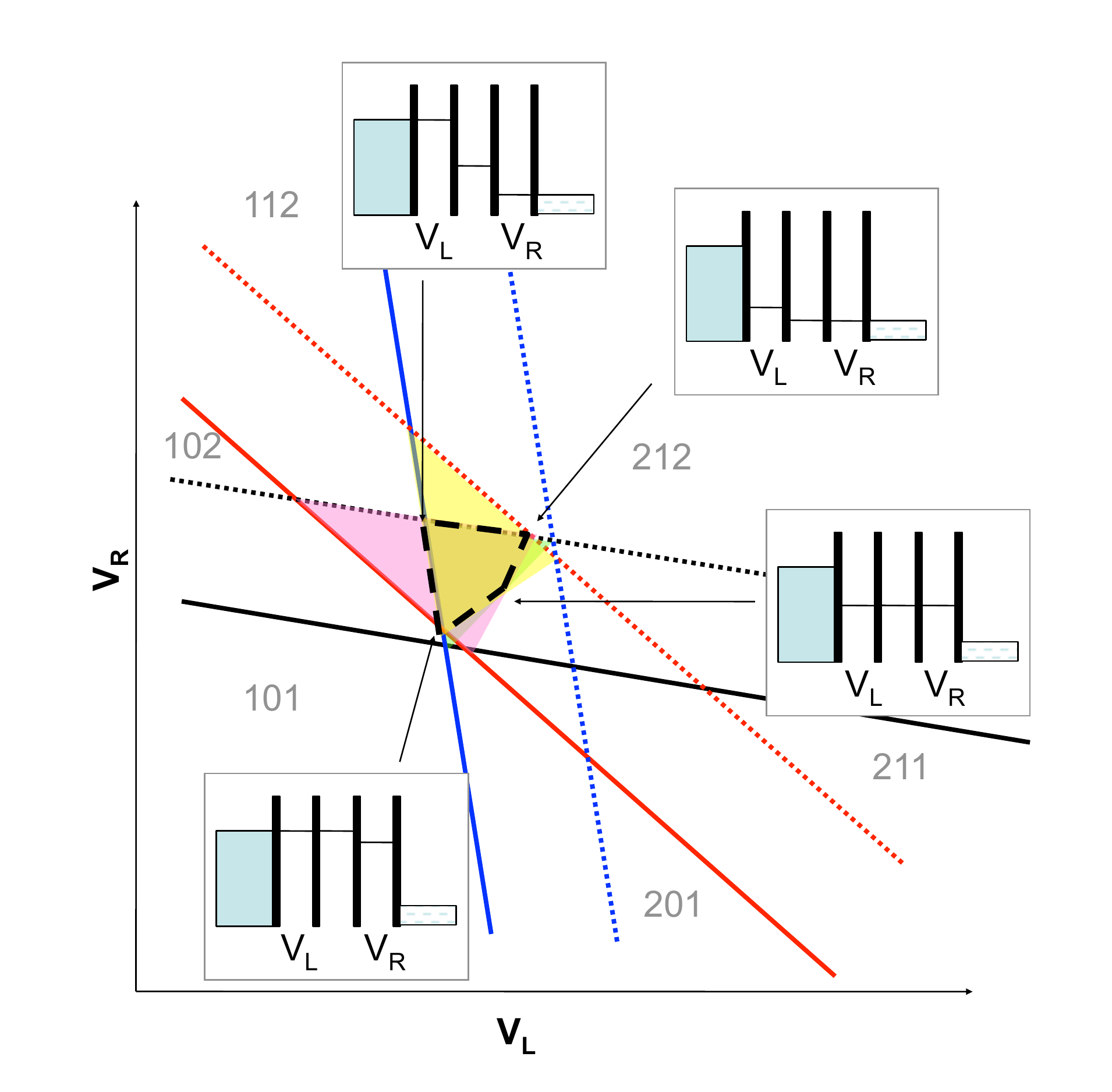}
\end{center}
\caption{
{\bf Shape of transport regions and energy diagrams in the presence of a drain-source bias across the TQD near QPs.} The chemical potential for the left (centre) [right] dots are shown as blue (red) [black] lines. The solid (dotted) lines correspond to an alignment of a dot chemical potential with the left (right) lead. The capacitive couplings and tunnel couplings are neglected for this diagram. We assume current could flow between any given pairs of dots in triangular regions as shown by the partly overlapping yellow, pink, and green areas. The size of the triangles is proportional to the bias.  The intersection of the three coloured triangles is where current is expected to actually flow, and this region is quadrangular for the case drawn here (black dashed quadrangle). The energy diagrams for the TQD at each of the four vertices are also drawn.}
\label{fig:tqdbsb_quadrangle}
\end{figure}

\begin{figure}
\begin{center}
\includegraphics[width=3.8in,clip]{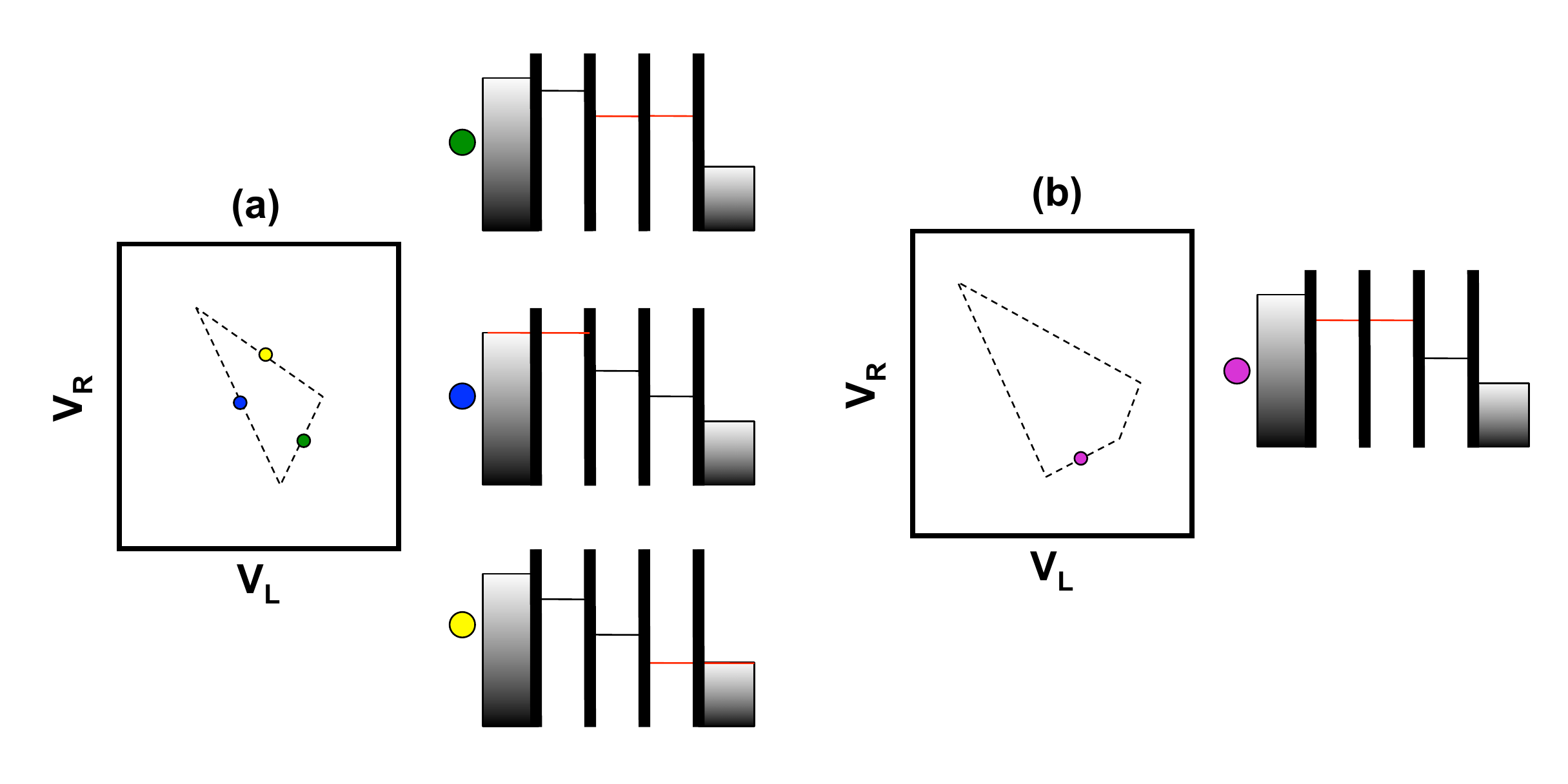}
\end{center}
\caption{
{\bf Quantum dot chemical potentials.} {\bf a}, Diagrams showing the chemical potentials at specific places along the edges of the transport triangle where transport through the triple quantum dot occurs. {\bf b}, Under certain conditions, the transport regions can be bounded by four lines of different slopes, including the resonance between the left and central dots.}
\label{fig:tqdbsb_triangles}
\end{figure}

\section*{S3. Identifying the resonances in experimental transport data}
The identification of the resonance lines from experimental transport data is crucial to the understanding of the bipolar spin blockade and the coherent superposition mechanism but surprisingly simple to achieve. Here we explain the procedure that is followed in order to determine what dots are in resonance along any given line inside the boundary of the transport diagram.

The first step is to measure the stability diagram using charge detection at zero bias where three addition lines cross in order to determine the slopes of the electron addition lines for each dot as well as the slopes of the charge transfer lines where an electron is transferred between dots~\cite{grangerPRB10}.

Once a bias is applied across the TQD, the chemical potential for each dot can be aligned separately with either the source or the drain chemical potential. An idealized diagram in the $V_\text{L}-V_\text{R}$ plane is shown in Fig.~\ref{fig:tqdbsb_quadrangle}, where the addition lines corresponding to the addition of an electron from the left lead are represented as solid lines and those corresponding to the addition of an electron from the right lead are represented as dotted lines. The capacitive couplings and tunnel couplings are neglected here, i.e. charge transfer lines are not drawn for simplicity. We make a simple assumption that the presence of the bias would allow current to flow between any given pairs of dots in a triangular area of the stability diagram bounded by the following lines: the addition line for the leftmost dot of the pair from the left lead; the addition line for the rightmost dot of the pair from the right lead; and a line parallel to the charge tran
 sfer line of this pair of dots (not shown). The distance from this line to the intersection of the two aforementioned lines is proportional to the drain-source bias.  In Fig.~\ref{fig:tqdbsb_quadrangle}, we draw the three triangles corresponding to the three pairs of dots. The actual measurement would show transport within the region of the transport diagram that corresponds to the intersection of these three triangles. In the particular case depicted in Fig.~\ref{fig:tqdbsb_quadrangle}, these regions form a quadrangle. The energy diagrams for the TQD at each of the four vertices are drawn in Fig.~\ref{fig:tqdbsb_quadrangle}.

The resonance lines in the transport stability diagram inside the transport ``triangle'' are parallel to the respective charge transfer lines from the zero bias charge detection stability diagram. The slopes of the charge transfer lines correspond to the situation when two particular dots are on resonance. We use these slopes, therefore, to identify the resonances directly. They are also consistent with the theoretical calculations.

We note that it is more usual to observe the triangle than quadrangle. This is because the point where dots L, C, and R are all in resonance often occurs outside the transport window once a bias is applied. Indeed, even though the TQD is tuned in order to see the QPs at zero bias, the application of a bias across it shifts the chemical potentials in such a way that the gate voltage configuration that restores the perfect alignement of the three chemical potentials may fall outside the transport regions. As this point is responsible for the fourth vertex in Fig.~\ref{fig:tqdbsb_quadrangle}, it is often lost. One then measures a transport triangle instead of a quadrangle (see Fig.~\ref{fig:tqdbsb_triangles}).


\begin{thebibliography}{00}

\bibitem{pettaSC05}
Petta, J. R.  {\it et al.} Coherent Manipulation of Coupled Electron Spins in Semiconductor Quantum Dots. {\it Science} {\bf 309,} 2180-2184 (2005).


\bibitem{hansonPRL07}
Hanson, R. \& Burkard, G. Universal Set of Quantum Gates for Double-Dot Spin Qubits with Fixed Interdot Coupling. {\it Phys. Rev. Lett.} {\bf 98,} 050502 (2007).

\bibitem{onoSC02}
Ono, K., Austing, D. G., Tokura, Y. \& Tarucha, S. Current Rectification by Pauli Exclusion in a Weakly Coupled Double Quantum Dot System. {\it Science} {\bf 297,} 1313-1317 (2002).

\bibitem{gaudreauNAPHYS12}
Gaudreau, L. {\it et al.} Coherent control of three-spin states in a triple quantum dot. {\it Nature Physics} {\bf 8,} 54-58 (2012).

\bibitem{hsieh2012}
Hsieh, C.-Y., Shim, Y.-P. \& Hawrylak, P. Theory of electronic properties and quantum spin blockade in a gated linear triple quantum dot with one electron spin each. \textit{Phys. Rev. B} \textbf{85,} 085309, 2012.
	
\bibitem{arimondoCPT} 
Arimondo, E. {\it Coherent Population Trapping in Laser Spectroscopy Ch. V} (E. Wolf Progress in Optics XXXV, 1996).

\bibitem{brandesPRL00}
Brandes, T. \& Renzoni, F. Current Switch by Coherent Trapping of Electrons in Quantum Dots. {\it Phys. Rev. Lett.} {\bf 85,} 4148-4151 (2000).

\bibitem{emaryEPL06}
Michaelis, B., Emary, C. \& Beenakker, C. W. J. All-electronic coherent population trapping  in quantum dots. {\it Europhys. Lett.} {\bf 73,} 677-683 (2006).

\bibitem{ratnerJPhysChem90}
Ratner, M. A. Bridge-assisted electron transfer: effective electronic coupling. {\it J. Phys. Chem.} {\bf 94,} 4877-4883 (1990).

\bibitem{Greentree2004} Greentree, A.D.,  Cole, J.H.,  Hamilton, A.R. \& Hollenberg, L.C.L. Coherent electronic transfer in quantum dot systems using adiabatic passage. \textit{Phys. Rev. B} \textbf{70,} 235317 (2004).

\bibitem{Johnson2005}  Johnson, A. C.,  Petta, J. R.,   Marcus, C. M.,  Hanson, M. P. \& Gossard, A. C. Single-triplet spin blockade and charge sensing in a few-electron double quantum dot. \textit{Phys. Rev. B} \textbf{72,} 165308 (2005).

\bibitem{koppensSC05}
Koppens, F. H. L. {\it et al.} Control and Detection of Singlet-Triplet Mixing in a Random Nuclear Field. {\it Science} {\bf 309,} 1346-1350 (2005).

\bibitem{grangerPRB10}
Granger, G. {\it et al.} Three-dimensional transport diagram of a triple quantum dot. {\it Phys. Rev. B} {\bf 82,} 075304 (2010).

\bibitem{maria}
Busl, M., S\'anchez, R. \& Platero, G. Control of spin blockade by ac magnetic fields in triple quantum dots. {\it Phys. Rev. B} {\bf 81}, 121306(R) (2010).

\bibitem{amaha2012} Amaha, S.,  Hatano,  T., Tamura, H., Teraoka, S., Kubo, T.,  Tokura, Y., Austing, D. G.,  and Tarucha, S. Resonance-hybrid states in a triple quantum dot. {\it Phys. Rev. B} {\bf 85,} 081301(R) (2012).

\bibitem{blum}
Blum, K. Chapter 8 in {\it Density Matrix, Theory and Applications, second ed.} (Plenum Press, New York, London, 1996)

\bibitem{gaudreauPRL06}
Gaudreau, L. {\it et al.} Stability Diagram of a Few-Electron Triple Dot. {\it Phys. Rev. Lett.} {\bf 97}, 036807 (2006).



\end{thebibliography}
\end{document}